\documentclass{emulateapj}

\usepackage{amsmath,amssymb}
\usepackage{natbib}
\usepackage{epsfig,graphics}
\usepackage{epstopdf} 

\shorttitle{Non-Detection of L-band Emission from HD189733b}
\shortauthors{Mandell, Deming, Blake, Knutson, Mumma, Villanueva, \& Salyk}

\begin{document}
\title{Non-Detection of L-band Line Emission from the Exo-Planet HD189733b} 
\author{Avi M. Mandell\altaffilmark{1,6}, L. Drake Deming\altaffilmark{1}, Geoffrey A. Blake\altaffilmark{2}, Heather A. Knutson\altaffilmark{3}, Michael J. Mumma\altaffilmark{1}, Geronimo L. Villanueva\altaffilmark{1,4}, \& Colette Salyk\altaffilmark{5}} 
\altaffiltext{1}{Solar System Exploration Division, NASA Goddard Space Flight Center, Greenbelt, MD 20771, USA}
\altaffiltext{2}{California Institute of Technology, Division of Geological and Planetary Sciences, MS 150-21, Pasadena, CA 91125, USA}
\altaffiltext{3}{University of California Berkeley, Department of Astronomy, 601 Campbell Hall, Berkeley, CA 94720 USA}
\altaffiltext{4}{Catholic University of America, Department of Physics, Washington, DC 20064, USA}
\altaffiltext{5}{The University of Texas at Austin, Department of Astronomy, 1 University Station C1400, Austin, Texas 78712, USA}
\altaffiltext{6}{Corresponding Email:  Avi.Mandell@nasa.gov}

\begin{abstract}
We attempt to confirm bright non-LTE emission from the exoplanet HD\,189733b at 3.25\,$\mu$m, as recently reported by \citet{swain2010p637} based on observations at low spectral resolving power ($\lambda/\delta\lambda \approx 30$).  Non-LTE emission lines from gas in an exoplanet atmosphere will not be significantly broadened by collisions, so the measured emission intensity per resolution element must be substantially brighter when observed at high spectral resolving power. We observed the planet before, during, and after a secondary eclipse event at a resolving power $\lambda/\delta\lambda = 27,000$ using the NIRSPEC spectrometer on the Keck II telescope. Our spectra cover a spectral window near the peak found by Swain et al., and we compare emission cases that could account for the magnitude and wavelength dependence of the Swain et al. result with our final spectral residuals.  To model the expected line emission, we use a general non-equilibrium formulation to synthesize emission features from all plausible molecules that emit in this spectral region.  In every case, we detect no line emission to a high degree of confidence. After considering possible explanations for the Swain et  al. results and the disparity with our own data, we conclude that an astrophysical source for the putative non-LTE emission is unlikely.  We note that the wavelength dependence of the signal seen by Swain et al. closely matches the $2\nu_{2}$ band of water vapor at 300K, and we suggest that an imperfect correction for telluric water is the source of the feature claimed by Swain et al.
\end{abstract}

\keywords{astrobiology -- infrared: planetary systems -- planets and satellites: individual (HD189733b) -- radiative transfer -- techniques: spectroscopic}

\section{Introduction}
Characterization of exoplanet atmospheres is a highly active field of research, especially using transit and eclipse techniques. Space-borne detections of transiting exoplanet atmospheres have become commonplace, both in the infrared (e.g., \citealt{charbonneau2005p523, Deming2005p740, Harrington2007p691, charbonneau2008p1341, swain2008p482, knutson2009p822, Pont2008p109, Pont2009pL6, grillmair2008p767, fressin2010p374, gillon2010p3}) and optical spectral regions (e.g., \citealt{charbonneau2002p377, snellen2009p543, Snellen2010p76}).  The space-borne data have been extensively exploited to investigate the temperature structure and dynamics of hot exoplanet atmospheres (e.g., \citealt{burrows2007pL171, Burrows2008p1436, fortney2008p1419, showman2009p564, fortney2010p1396}; see \citealt{seager2010p631} for a recent review). 

Near-infrared photometry of the hottest exoplanets is now possible using ground-based observatories (e.g., \citealt{Gillon2009p359, alonso2010p1481, deMooij2009pL35}), and spectroscopy of strong atomic lines has also been successful from the ground (e.g., \citealt{redfield2008pL87, Snellen2008p357}).  Since a new generation of extremely large ground-based telescopes is now being planned \citep{Hook2009p225}, ground-based spectroscopy of molecular features in exoplanet atmospheres may eventually become possible, and would provide exceptional diagnostic power. So far, ground-based attempts to detect near-IR molecular features in exoplanet atmospheres with a range of spectral resolving powers have for the most part been unsuccessful (e.g., \citealt{Wiedemann2001p1068, Brown2002p826, Richardson2003p581, Richardson2003p1053, Barnes2007p473, Barnes2008p1258,Barnes2010p445}), though high-resolution spectroscopy has recently produced a successful detection of CO at K-band wavelengths \citep{Snellen2010p1049}. 

Additionally, a startling ground-based detection of molecular spectral features from the atmosphere of the exoplanet HD\,189733b at moderate spectral resolution was announced by Swain et al. (2010; hereafter S10). S10 analyzed observations of HD\,189733 taken with the SpeX instrument on the Infrared Telescope Facility, focusing on two wavelength regions: $2.0 - 2.4\,\mu$m (K band) and $3.1 - 4.1\,\mu$m (L band).  Their results show a rising spectral slope at 2.2\,$\mu$m, which they attribute to CO$_2$ absorption, and a bright peak at 3.25\,$\mu$m. The shape of the spectrum at K-band wavelengths matched previous results at these wavelengths taken with the Hubble Space Telescope \citep{swain2009pL114}, which the authors construed as support for their ground-based detection in the L band.

S10 interpret the flux peak in the L-band portion of their data as evidence of bright non-LTE (NLTE) emission from radiatively-excited methane. The wavelength region for the claimed emission roughly corresponds with the R-branch region of CH$_4$ ($3.1-3.34\,\mu$m), but no corresponding feature is seen in the region of the methane P-branch ($3.35-3.5\,\mu$m); S10 note this disparity but offer no theoretical explanation for it. The L-band emission claimed by S10 is also remarkable for having high intensity at low spectral resolution. The wavelength extent of each spectral bin analyzed by S10 is 0.1\,$\mu$m, ($\lambda / \delta\lambda \sim 30$), and they find a planet-to-star contrast ratio of approximately 0.9\% in their brightest bin at 3.25\,$\mu$m. This result is even more surprising considering that NLTE emission lines are unlikely to be collisionally broadened, since the pressures needed to achieve significant line broadening would produce high collision rates that would drive the level populations to thermal equilibrium. Therefore any NLTE emission lines will have maximum intrinsic widths comparable to the Doppler velocities of the emitting species, and when observed at low resolving power the apparent flux densities (ergs s$^{-1}$ cm$^{-2}$ Hz$^{-1}$) of emission lines will be greatly reduced because their intrinsic line widths are unresolved. This motivated our attempt to confirm the S10 results using data with a much higher spectral resolving power, obtained with the Near-IR Spectrograph (NIRSPEC; \citealt{mclean1998p566}) on the 10-m Keck II telescope . The peak intensities of NLTE emission lines will be much greater when they are observed at high spectral resolving power, and S10's claim of bright emission at low spectral resolution implies that observations at high spectral resolution and sensitivity should easily produce a detection.

A description of our search for absorption signatures of the exoplanet in our data set will be presented in a future manuscript; the expected line depths based on standard atmospheric modeling are approximately $0.01-0.1$\% of the stellar continuum \citep{Burrows2008p1436}, and a comprehensive analysis at that level is beyond the scope of this study. In this paper we limit our analysis and discussion to whether or not we can confirm the bright emission signal announced by S10. We present our observing and data analysis procedure in Section \S\ref{obs}, predict the expected emission features using excitation models of molecular emission at different rotational temperatures and compare the expected signal to our results in Section \S\ref{res}, and conclude with a discussion of the potential explanations for the differences between our results and those of S10 in Section \S\ref{disc}.

\section{Observations and Data Analysis}
\label{obs} We acquired spectra at $\lambda/\Delta\lambda \approx $ 27,000 with NIRSPEC without AO on UT July 13, 2009 using the KL filter with a setting covering portions of the wavelength range between 3.27 and 4.0\,$\mu$m.  Observing conditions were optimal, with low water vapor and clear skies.  We observed HD\,189733 before, during and after a secondary eclipse of the planet, for a total integration time of 100 minutes between UT 10:00 and UT 14:00. A bright B-type comparison star (HR\,8634) was also observed immediately after the science target. We nodded the telescope 12 arcsec in an ABBA sequence, with 60-second integrations per beam for both stars.  In total, we obtained 48 echelle spectra of HD\,189733 during eclipse, 52 spectra out of eclipse, and 40 spectra of HR\,8634.

Since our spectra were acquired prior to the publication of the S10 results, the wavelength ranges of our echelle grating orders are not identical to the S10 work.  Fortunately, our data include a spectral range from $3.27-3.31\,\mu$m that overlaps the brightest bin in the emission feature claimed by S10. The wavelength structure of molecular bands is known unequivocally from quantum mechanics, with only the level populations affecting the intensity of lines in different spectral channels.  Our modeling (see Section \S\ref{res}) indicates that molecules emitting strongly in the S10 3.25\,$\mu$m bin must also emit significantly in our $3.27-3.31\,\mu$m region, and we test the S10 results on that basis.

We utilized custom data reduction algorithms, previously used to detect new molecular emission features from warm gas in circumstellar disks \citep{mandell2008pL25}, to extract and process spectra for each echelle order in each ABBA set. We reduced the initial 2D spectral-spatial images to 1D spectra after first correcting for the slope of the beam due to cross-dispersion and subtracting A- and B-beam images to remove the contribution from telluric radiance. We identified bad pixels and cosmic ray hits in each raw pixel column by comparing the beam profile to an average beam profile for nearby columns, allowing us to identify and replace single-pixel events without removing any enhancements due to emission or absorption features.  

\begin{figure*}[htb]
\centering{
\includegraphics[width=170mm]{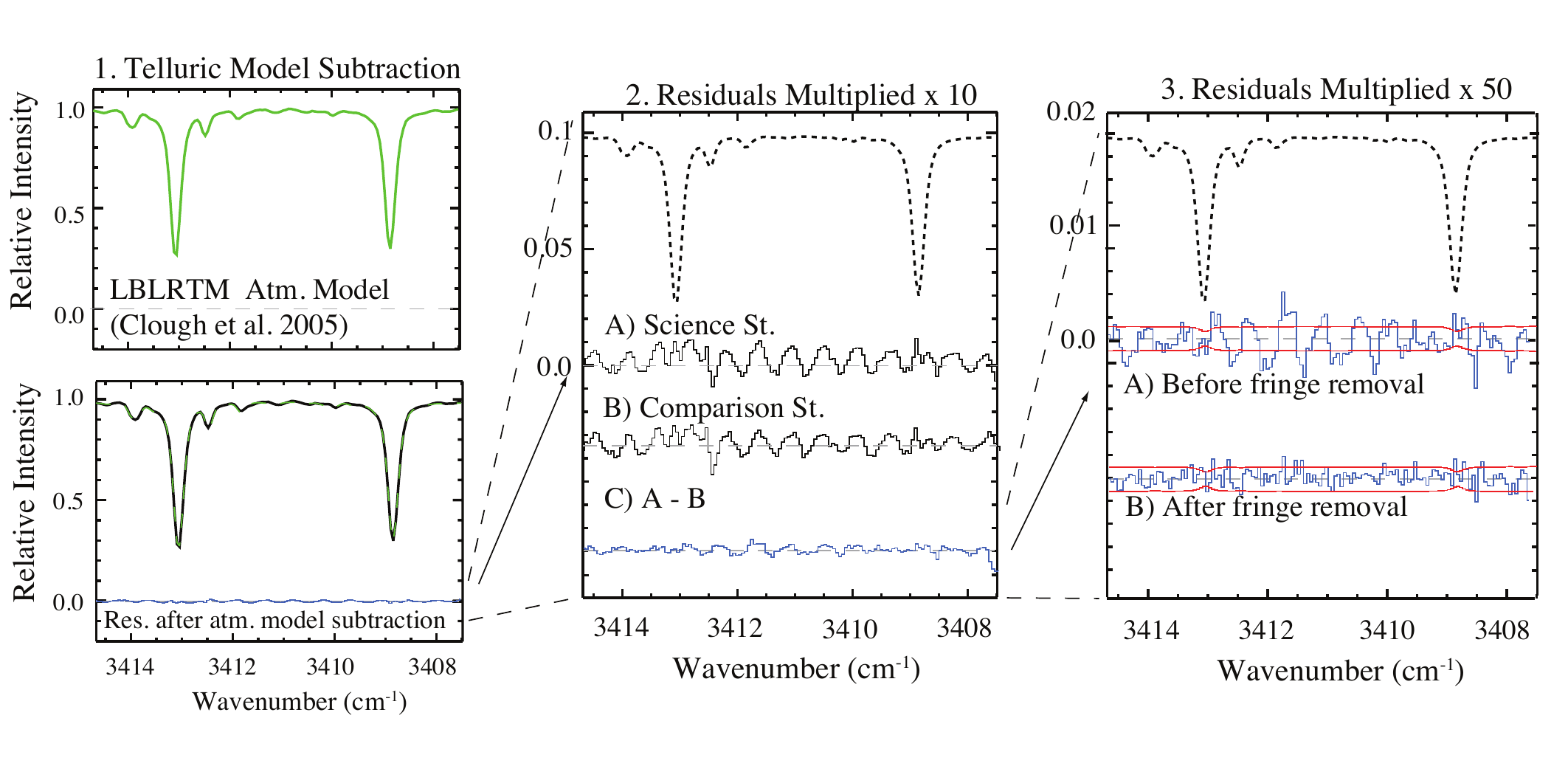}
\caption[scale=1.0]{\label{reduction} Demonstration of our data analysis procedure. Synthetic terrestrial atmospheric models are fitted separately to both the science and comparison stars and subtracted (1); then the two sets of residuals are differenced to remove second-order instrumental or atmospheric features (2).  Remaining fringing is removed using a Fourier filter (3).  The S/N in the final residuals is $\sim1000$. Stellar features are removed by observing the star during the eclipse of the planet and using the data to create a template of the stellar spectrum which is then subtracted from the out-of-eclipse
data.  }
}
\end{figure*}

We corrected for changing airmass and telluric atmospheric conditions to high accuracy using a two-step process: 1) fitting the data for both the science star (HD\,189733) and the comparison star (HR\,8634) with terrestrial spectral transmittance models synthesized with the LBLRTM atmospheric code \citep{clough2005p233} and subtracting the models to obtain spectral residuals for each star, and 2) differencing the residuals of the two stars. The subtraction of the telluric model compensates for the effects of changing airmass and atmospheric variability, and the differencing of the residuals removes remnant fringes and other instrumental artifacts, as well as minor errors in the telluric model such as imprecise pressure broadening, weak features missing from the line list, and inaccurate isotopic ratios.  For our LBLRTM models we utilized a standard tropical temperature profile and line parameters from the HITRAN2008 molecular database with updates from 2009 \citep{rothman2009p533}, and we fitted for the abundance of three key atmospheric constituents (H$_2$O, CH$_4$, and O$_3$) and a scaling factor for the temperature of the troposphere. Atmospheric models were fitted for wavelength sub-sections of each AB set, with the fully-resolved model convolved with the local instrumental resolving power derived for each sub-section to compensate for variability in the effective spectral resolving power due to the position of the spectrum on the detector. Additionally, using a telluric model provided us with an extremely well-calibrated wavelength solution for each sub-section ($\Delta\lambda/\lambda\sim10^{-6}$). This process achieved results corresponding to S/N $\sim 300$ on the original stellar continuum for each AB set, and an rms noise only slightly larger (by 20\%)  than that expected from the photon statistics (see Figure \ref{reduction}).

\begin{figure}[htb]
\centering{
\includegraphics[width=85mm]{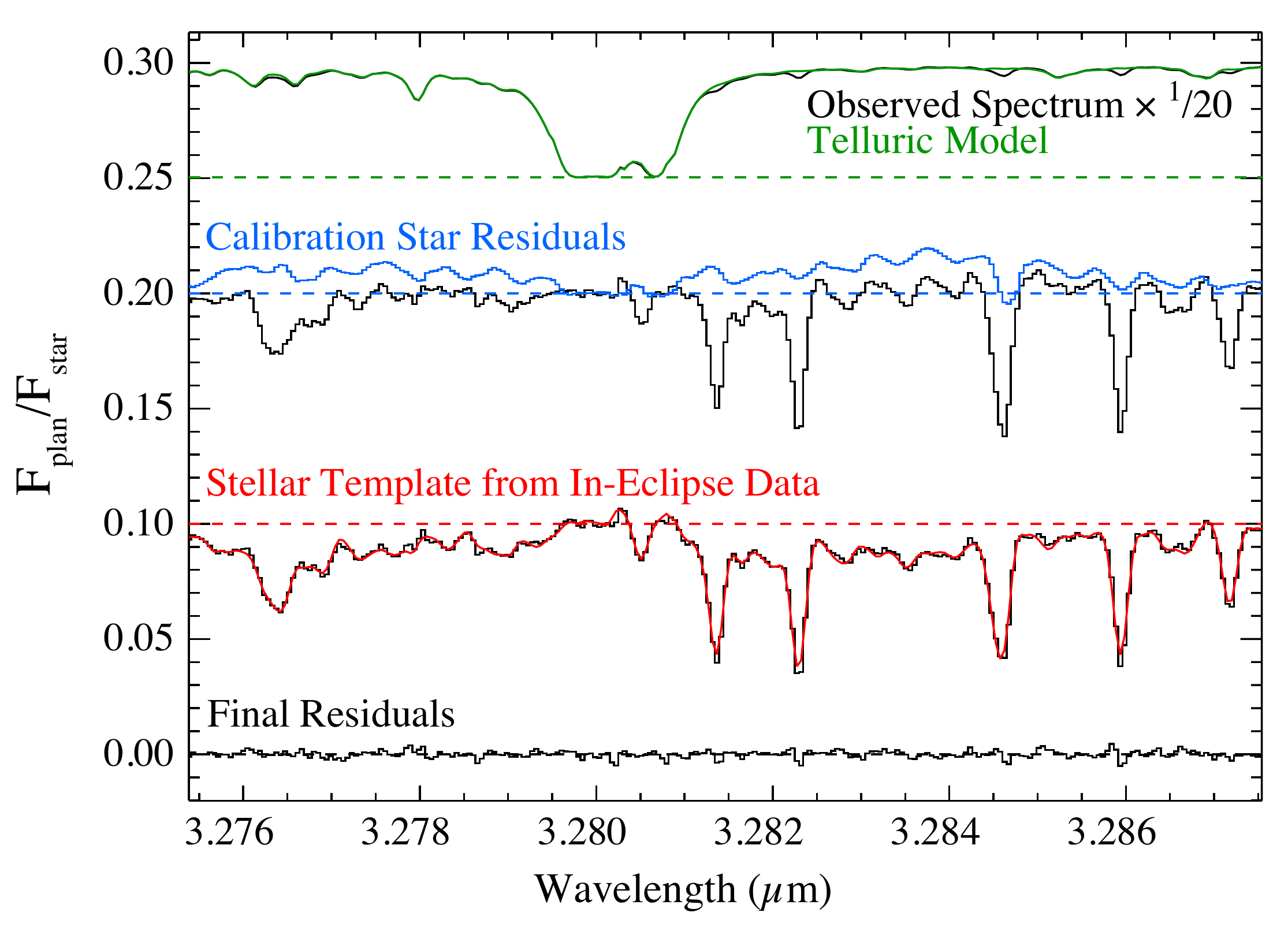}
\caption{\label{example} Sample spectral region for a single AB data set for HD\,189733.  A telluric model (green) is fitted to the observed spectrum and subtracted to produce an initial set of residuals.  We then subtract residuals from a similarly processed calibration star (blue) to remove instrumental errors and systematic deviations from the telluric model. We then bin the in-eclipse data to create a stellar template (red) which we subtract from the out-of-eclipse data to achieve our final residuals.  Low-amplitude deviations remain in the final residuals, primarily due to changes in the instrumental resolving power, but these features are eliminated when multiple sets are combined due to the changing radial velocity shift.}
}
\end{figure}

Stellar absorption features were removed by using the results for the in-eclipse data as a stellar template, with each AB set shifted to correct for changes in the heliocentric velocity and the stellar radial velocity variation due to the influence of the planet.  We then subtracted this stellar template from each AB set in the out-of-eclipse data after shifting for changes in the radial velocity for that time interval, leaving final residuals for each out-of-eclipse AB set with only the signal from the planet (see Figure \ref{example}).

We combined all the out-of-eclipse residuals after shifting each set to correct for the radial velocity of the planet around the star, with velocity shifts calculated based on the ephemeris from \citet{knutson2009p822}.  The radial velocity of the HD 189733 system relative to the telescope (accounting for Earth's orbit and rotation) was -8 km/s, but over our observing window the exoplanet's orbit led to velocity shifts over 60 km/s.  Because the data must be shifted in velocity space to account for the shift of the planet's spectrum, sections of the spectrum with high transmittance may be shifted to line up with spectral regions of low transmittance; this benefits us by moderating extinction of lines with strong telluric counterparts.  However, to avoid degrading our detectable signal by combining low-transmittance data with high-transmittance data, we weighted the value for each shifted spectral channel by the stochastic-noise SNR (based on detected photons and transmittance at the pre-shifted wavelength).

Finally, we applied a high-pass filter to the residuals for each set to remove variations in the continuum due to movement of the star on the detector over the course of the night by creating a smoothed version of the spectrum using a boxcar average with a window sufficiently large that any narrow features would be significantly reduced ($\sim20$ resolving elements), and then subtracted the filtered spectrum to achieve a final rms scatter of $\sim$0.001.  We evaluated the loss of narrow features due to the filtering by injecting synthetic emission lines and determined that the high pass filter preserved 96\% of their amplitudes.  Flux calibration was performed by normalizing the observed continuum flux to the predicted flux based on the K-band stellar magnitude \citep{cutri2003p?} and effective temperature \citep{vanbelle2009p1085}.

\section{A Search for Molecular Emission Features at High Resolution}
\label{res} In order to understand the origin of the excess emission signal detected by S10 between 3.0 and 3.5\,$\mu$m, we must first understand the potential contributing emitting species and their spectral characteristics. S10 suggest that methane emission at 3.3\,$\mu$m provides the best fit to their data, but there are clearly other species with transitions in this spectral region that may also be sufficiently abundant to produce detectable emission in the planet's atmosphere.  In order to constrain the potential molecular species that could be producing emission, we generated models of line emission for a range of possible molecular species and compared the resulting predictions to our data.  The goal of these models is not to explain the S10 results through a prediction of emission from a specific mechanism; rather, we only want to mimic the location and magnitude of any emission lines that might arise from a variety of molecules assuming the most basic physical requirements, under a range of excitation conditions, and then scale this emission based on the S10 results.

The dominant gas-phase molecular features in the L band are ro-vibrational spectral lines, in which a molecule transitions from a specific higher-energy rotational and vibrational configuration to a lower-energy rotational and vibrational state. The population of molecules in any specific rotational and vibrational state is determined by the transition rates between vibrational energy states as well as the transition rates between rotational energy states. The fundamental ratios between the strengths of different transitions for a gas in local thermodynamic equilibrium (LTE) are set by the Boltzmann distribution and are based on the temperature of the gas and the probability for each individual transition.  Any NLTE mechanisms such as fluorescence (radiative pumping by incident photons) or charged-particle interactions will increase the excitation rates and subsequent emission intensities; the level populations are therefore a function of both the underlying equilibrium distribution as well as any NLTE excitation processes. In particular, fluorescent excitation of irradiated gas may result in an increased strength for higher-energy vibrational bands whose upper state populations are pumped by high-energy photons from a nearby radiation source (e.g., the central star).

For comparisons with the data, we do not want to assume a specific physical excitation mechanism {\it a priori} (we discuss the potential for different excitation mechanisms in more detail in Section \S\ref{disc}), but we require that the basic rotational and vibrational structure of each molecule's spectrum is maintained. Since transitions between rotational energy states require much less energy than vibrational transitions, we first examined whether rotational transitions for energy states in the L band would be dominated by LTE or NLTE excitation. Using excitation rates from the HITRAN2008 molecular database with updates from 2009 \citep{rothman2009p533}, and collisional rate coefficients for rotational transitions of our candidate molecules with molecular hydrogen (10$^{-11}$ cm$^3$/s, \citealp{faure2008p137,faure2008p257}) and free electrons ($10^{-6} - 10^{-9}$ cm$^3$/s, \citealp{faure2004p323}), rotational thermalization of the ground vibrational states is expected to be achieved at pressures greater than 0.01 millibars.  Only at very high altitudes, where total column densities are negligible and many molecular species would be photodissociated \citep{liang2003pL247}, does rotational equilibrium fail.  Therefore, the rotational structure for any ro-vibrational emission from the exoplanet will broadly resemble a Boltzmann distribution for the rotational population, thermalized to the local kinetic temperature, and any additional NLTE excitation will only affect the intensity of different vibrational energy states.

\begin{figure*}[htb]
\centering{
\includegraphics[width=170mm]{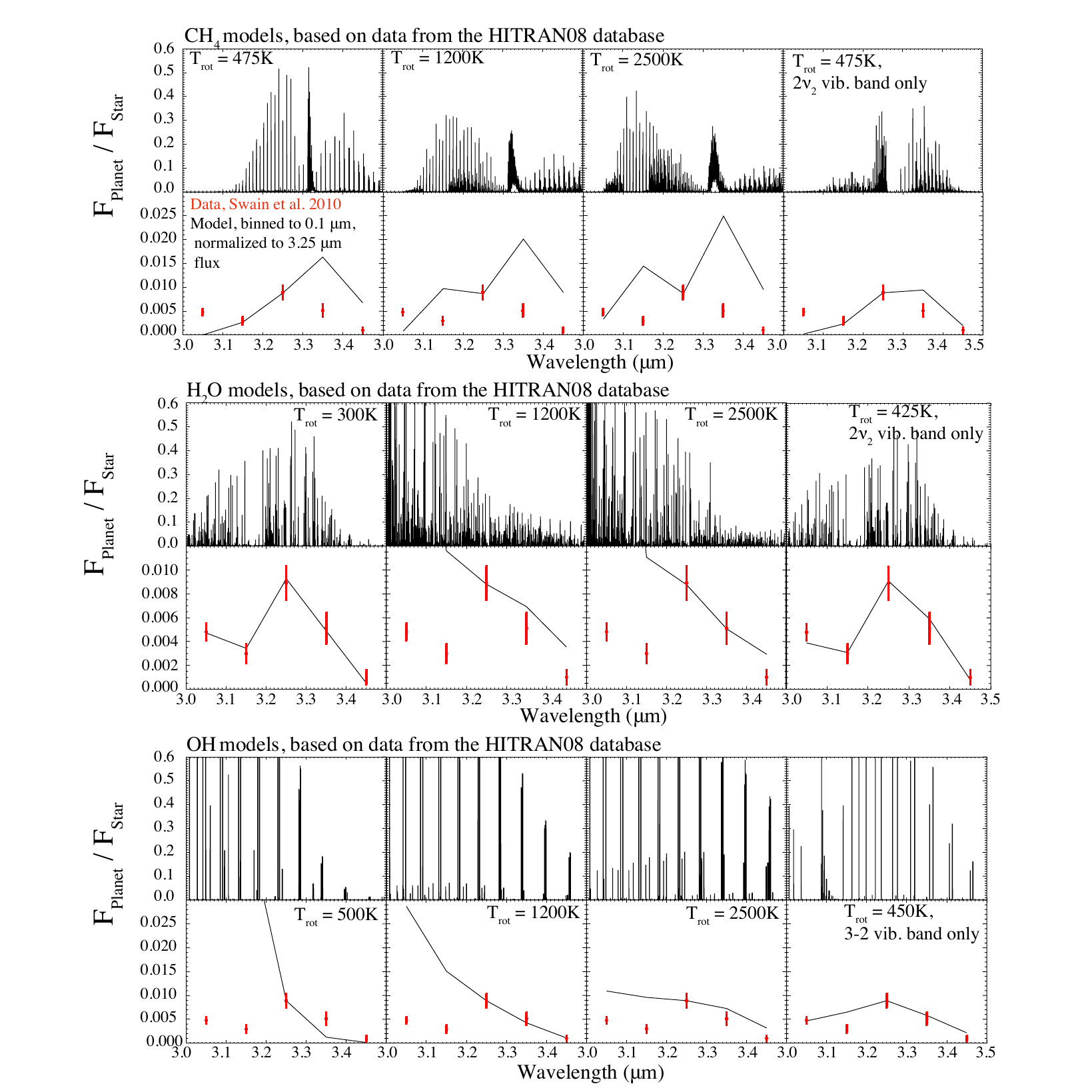}
\caption{\label{models} Excitation models for the three candidate molecular species that we compare to our data to search for expected emission features. For each molecule, we calculate three models using a single excitation temperature for the Boltzmann distribution of both the rotational levels and the vibrational levels.  In the fourth panel on the right, we plot the best-fit model using only a single vibrational level, in order to test the potential effects of strong radiative pumping of an upper vibrational level.  We scale each model to the 3.25\,$\mu$m planet-star contrast from \citet{swain2010p637}, and overplot their results in red. H$_2$O clearly provides the best fit. The single-vibrational-level models for CH$_4$ and OH also fit reasonably well, but the scaling required to reproduce the flux reported by \citet{swain2010p637} requires unphysical pumping efficiencies.}
}
\end{figure*}

We therefore calculate relative line strengths for ro-vibrational transitions for a candidate molecule using two bracketing cases: 1) we adopt the simplest case of a common rotational and vibrational excitation temperature (i.e., all vibrational bands are collisionally excited) from an optically thin gas, and 2) we postulate that emission from a single radiatively excited vibrational band dominates the spectrum. This formulation is quite general, and allows us to test the results of exciting a specific vibrational band without prior assumptions about the relative vibrational level populations. We first calculated rotational and vibrational levels populated in proportion to a single characteristic excitation temperature that we varied over a wide range (from 100K to 10,000K, stepping by 0.2 in log$_{10}$ space); examples are compared with the S10 results in Figure \ref{models}. We then calculated models with only a single vibrational level populated. For each of these cases, we calculate the emission spectrum of each given molecule, and divide that spectrum by a \citet{Kurucz1993p?} model stellar spectrum that matches the star (T$_{eff}=5000$K, log(g)$=4.5$, Z$=0.0$).  We adjust the absolute level populations of the emitting molecule so that the resultant contrast (planet/star) equals the observed S10 value at 3.25\,$\mu$m, and we use a $\chi^2$ fit to the five data points between 3.05 and 3.45\,$\mu$m from S10 to determine the best-fit rotational temperature for both the single-temperature models and the single-vibrational-level models.  We then examine the implied molecular spectrum in our adjacent wavelengths to determine whether our spectra are consistent with that molecule at that excitation temperature.

We determined that the species with transitions in this region, and the potential for a reasonable number of emitting molecules, are H$_2$O, CH$_4$, NH$_3$, and OH; modeling quickly shows that NH$_3$ peaks below 3.25\,$\mu$m for all temperatures and vibrational levels. In Figure \ref{models}, we plot the best-fit single-temperature models for the other three molecules (475K for CH$_4$, 300K for H$_2$O and 500K for OH), as well as two additional single-temperature models covering a large range in excitation temperature. We also plot the best-fit single-vibrational-band model for each molecule.  In the lower panels, the models are binned in the same spectral channels as S10 and scaled to match their data point at 3.25\,$\mu$m (their bin overlapping our measured wavelength region).

It is clear from our modeling that none of the single-temperature models for CH$_4$ and OH fit the S10 results.  The single-vibrational-level models allow a reasonably good fit, though each model has at least one major deviation from the S10 data. Additionally, the scaling factors required to fit the CH$_4$ and OH single-vibrational-level models to the data (10$^2$ and 10$^9$ respectively) require a level of radiative pumping that would be unattainable based on the stellar flux, as we explain below when considering possible fluorescence (Section \S\ref{fluor}).  There may be an intermediate model that fits the S10 data with even more precision, with a number of vibrational bands at a range of intensities, but we are mostly interested in the prediction for emission features in our spectrum and the bracketing cases considered here provide the full range of possible features; a full NLTE model is beyond the scope of this paper. Interestingly, the models for H$_2$O are able to fit the data very well ($\chi^2 \sim 1$) using low-temperature, single-excitation-temperature models (the single-vibrational-level model fits as well, due to the dominance of the single $2\nu_2$ vibrational band in the spectral region).  We discuss the potential implications of this result in Section \S\ref{cont}.

\begin{figure*}[htb]
\centering{
\includegraphics[width=170mm]{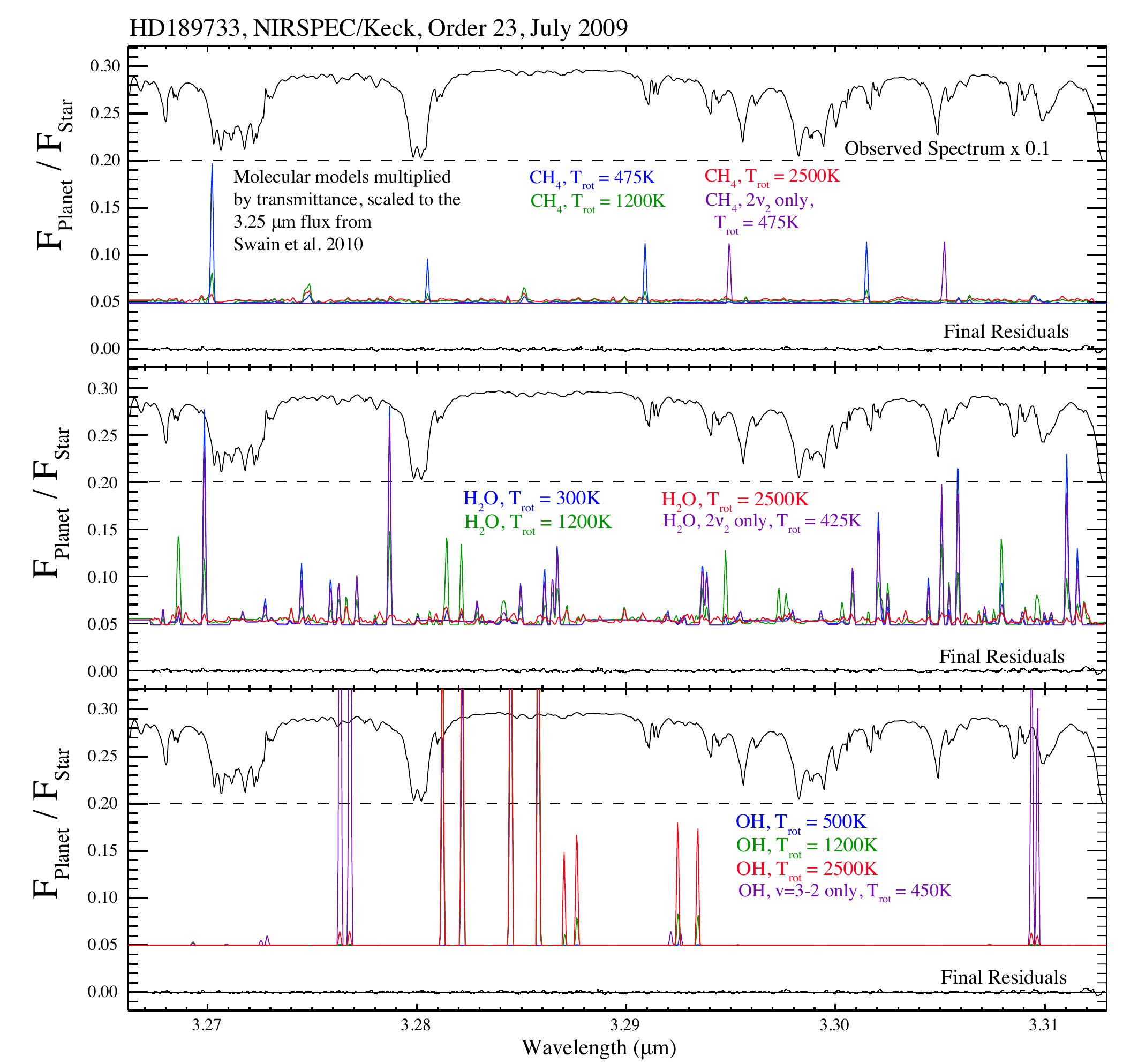}
\caption{\label{data} Results from our data reduction, with candidate molecular models using the CH4@Titan and HITEMP2010 line lists overplotted.  The upper black trace is the original data, and the bottom black trace are our residuals after removing telluric and stellar features, shifting the data to the correct radial velocity and combining it, and then removing large-scale gradients in the continuum with a high-pass filter (as described in \S\ref{obs}).  Models for CH$_4$ (top) and H$_2$O (bottom) with different excitation temperatures are plotted above. We detect no emission at any of the expected positions, with upper limits more than 10 times below the total intensity of the expected emission for each model.  }
}
\end{figure*}

In order to run a full grid of models for a range of temperatures and vibrational states for each molecule over a wide wavelength range, it was necessary to utilize a list of transitions that was both relatively complete and manageable in size. We primarily used data from the HITRAN2008 database for the full grid of models; however, we then compared our results to models computed with more complete line lists - data from the CH$_4$@Titan line database for methane \citep{ch4@titan2010p?} and data from the HITEMP2010  database \citep{Rothman2010p2139} for water and OH.  Differences between the low-resolution models based on HITRAN and the more complete models were virtually non-existent at lower temperatures, and while the flux for some increased up to 20\% for the highest temperatures, these differences are insignificant due to the already-poor fit between higher-temperature models and the S10 results as illustrated in Figure \ref{models} and the fact that our upper limits are significantly smaller than the expected signal. 

In Figure \ref{data}, we plot the scaled models based on the more complete data sets over our spectrum, which covers both high- and low-temperature methane and water transitions.  The models are multiplied by the telluric transmittance model (i.e., attenuated by extinction from the Earth's atmosphere) and filtered using the same high-pass filter used to remove the continuum in the data. None of the candidate molecular emission features that could account for the S10 result can be detected in our data.  Indeed, in every instance they can be ruled out unequivocally; in most cases the emission would be so bright that the features would be visible in the raw data.  We also performed the same data reduction procedure on three additional NIRSPEC orders, spanning wavelength ranges of $3.41-3.46\,\mu$m, $3.57-3.63\,\mu$m, and $3.75-3.81\,\mu$m; no emission was detected to the same sensitivity limits in any of the orders. The standard deviation of our observed residuals (lower trace in Figure \ref{data}) is 0.0011 at our observed resolution, and is even lower when averaged over a line width.  To see how firmly the various cases are rejected, consider the spectrum of methane excited at 1200K.  That spectrum requires a 3.5\% emission feature near 3.27\,$\mu$m that would be detected at more than $30\sigma$ significance in our data, but is not seen.  The other molecular cases illustrated in Figure \ref{data} are rejected at even higher levels of significance.  Moreover, as stated above, we have not restricted our search to merely the cases that are illustrated in Figure \ref{data}.  In no instance can we find a spectrum that accounts for the S10 results but is also allowed by our data.  We can only reconcile our data with the S10 result if we allow only the rotational transitions not covered by our spectral windows to contain flux, which we consider impossible  given the range of transitions with similar excitation energies between 3.0 and 3.5 \,$\mu$m as well as the high collisional cross-sections for changes in the rotational quantum number of water and methane \citep{faure2004p323,faure2008p137,faure2008p257}.

\section{Discussion}
\label{disc}
In this section we discuss possible explanations for the emission feature detected by S10, and evaluate whether these explanations are consistent with known physical emission mechanisms as well as the observational results from this study and previous studies of the HD189733 system.  We then address the possibility that the emission feature is not astrophysical in nature, but rather a data reduction artifact caused by incomplete removal of the signature of telluric water vapor.

\subsection{Fluorescent Pumping by the Stellar Continuum}
\label{fluor}The most well-understood NLTE process that can produce NIR line emission is fluorescence excited by radiation from the stellar continuum, either through direct pumping by incident radiation at resonant frequencies (resonant fluorescence) or through radiative cascades from upper energy levels (non-resonant fluorescence).  In our own Solar System, methane fluorescence is well known, for example from the atmospheres of Jupiter \citep{Drossart1999p169} and Titan \citep{brown2006p707}, and S10 claim methane fluorescence as the most likely source for their signal. 

Both resonant and non-resonant fluorescence pumped by photons from the stellar continuum can be rejected as the source of the S10 feature on both observational and theoretical grounds.  First, any emission would be present at a similar level in both 2007 and 2009, and we conclusively rule out the presence of bright narrow emission features that would be present in our data.  Furthermore, consideration of the energetics required to produce fluorescent emission shows that the emission feature claimed by S10 is so bright that it is impossible to attribute to fluorescence based on pumping by stellar radiation. 

For resonant fluorescence, any radiation scattered back to the observer by the planetary atmosphere cannot be more intense than the impinging stellar radiation; this provides an ``optically thick'' limit for an absorbing transition. The radii of the planet and star, and other system parameters, are well known from high-precision transit observations \citep{Winn2007p1828}.  Therefore, we know that the solid angle subtended by the planet is 2.1\% of the stellar solid angle as viewed from Earth. S10 find a contrast (planet flux divided by stellar flux) equal to 0.9\% in their brightest bin at 3.25\,$\mu$m.  If we start with the assumption that their feature is at least marginally resolved (i.e. their measured peak flux is not diluted over several resolving elements), that implies that the specific intensity (ergs cm$^{-2}$ sec$^{-1}$ sr$^{-1}$ Hz$^{-1}$) of the planet's emission at the top of its atmosphere is 43\% of the specific intensity of stellar radiation at the top of the stellar atmosphere. However, since the planet orbits at 8.4 stellar radii, the intensity of the stellar radiation field impinging on the planet is geometrically diluted by a factor of $8.4^2 = 70$. Hence, in order to produce the brightness of the S10 feature, the specific intensity of the emission from the planet would have to be 30 times greater than the impinging stellar radiation at 3.25\,$\mu$m to produce the peak flux measured at low resolution. 

If we assume that any emission is actually not resolved (as would be expected for NLTE emission from a low-density medium), the discrepancy becomes even more pronounced. For example, consider our molecular excitation model for methane at a 1200K rotational temperature, as described above.  This case produces a spectrally resolved emission line at 3.27\,$\mu$m that would peak at 3.5\% of the stellar continuum in our spectrum, but without extinction by the terrestrial counterpart would actually peak at 13.5\%.  Adjusting for the ratio of solid angles, the specific intensity in the core of this line is 6.5 times the stellar intensity at the top of the stellar atmosphere, at the same wavelength.  Since the stellar radiation field is diluted by a factor of 70 at the orbital distance of the planet, this specific planetary line would be 450 times brighter than the impinging stellar radiation field at that same wavelength. Resonant fluorescence is therefore incapable of producing the apparent flux observed by S10, whether the measured emission is fully resolved by S10 or not.

Although resonant fluoresence alone is not a feasible explanation for the S10 emission feature, we consider whether other fluorescence mechanisms could contribute significantly to the emission reported by S10. For example, stellar photons could in principle excite molecules to a higher vibrational level, and a subsequent radiative cascade or vibration-vibration (V-V) transfer by near-resonant collisions could produce emission in the 3.25\,$\mu$m bands. These processes are collectively known as non-resonant fluorescent pumping.  An example is the fluorescent emission seen in the 10\,$\mu$m carbon dioxide bands on Mars \citep{deming1983p356}.  In that case, radiative absorption at short wavelengths is followed by rapid V-V collisional transfer to the upper state of the fluorescent emission.  An analogous process might be energetically feasible for HD\,189733b, since the stellar pumping could occur at short wavelengths where a higher stellar photon flux is available, and the pumping could in principle involve multiple absorbing bands.

A detailed line-by-line non-resonant fluorescence model for a realistic planetary atmosphere is beyond the scope of this paper, but we can evaluate the degree to which non-resonant fluorescence can augment the flux from resonance fluorescence for the most relevant molecular candidate (methane). Non-resonant fluorescence of methane in planetary atmospheres has been discussed by \citet{Drossart1999p169} (Jupiter) and \citet{MartinTorres1998p1631} (Earth); in particular, \citet{MartinTorres1998p1631} evaluate the 19 strongest vibrational bands of methane in our own atmosphere, demonstrating that the contribution to bands at 3.3\,$\mu$m by radiative cascades from higher-energy states is insignificant.  We performed a similar calculation by using the CH4@Titan database (2.6 million transitions) to identify all the methane bands with upper-state energies greater than 3\,$\mu$m that could then cascade downward; in total we found 188 bands.  If we assume the absorbing bands are optically thin (appropriate for NLTE radiative pumping of higher-energy vibrational bands), we can use the stellar flux at each wavelength from a Kurucz stellar model combined with the Einstein B coefficient of absorption for each transition to calculate the total amount of stellar flux that could be absorbed per second by gas at 1200K and compare it to the total flux absorbed by all the tabulated bands between 3 and 3.5\,$\mu$m (a total of 52 bands).  This calculation does not include the branching ratios for each upper state, which determine how the absorbed energy is split between specific lower states, and therefore provides an upper limit to the fraction of absorbed energy that would actually cascade into the 3.3\,$\mu$m region. We find that non-resonance pumping by stellar photons and subsequent radiative cascades could augment the effect of resonance fluorescence by a maximum factor of 1.8.  This contribution is clearly insufficient to explain the energy difference between the impinging stellar radiation and the apparent emission detected by S10. We have similarly examined all our different molecular candidates scaled to the contrast claimed by S10 at 3.25\,$\mu$m, and every one of them can be ruled out using the same computational approach.

We therefore conclude that, even if the bright emission claimed by S10 is real, but variable so that it escapes our detection, it is too bright to be produced by fluorescent pumping from the stellar continuum.

\subsection{Time-Varying Emission from the Star or Planet}
Though the presence of gas-phase line emission can be definitively rejected in our data, the features detected by S10 could be attributed to a variable excitation or extinction mechanism that resulted in emission in 2007 and a lack of emission in 2009. Regardless of the specific mechanism needed to generate the emission claimed by S10, we can also consider more generally whether or not changes in the properties of the star or the planet might account for our non-detection of the putative signal. 

One possibility that might produce highly variable emission is stellar flares, which could produce molecular excitation at high altitudes by charged particle interactions or enhanced UV pumping.  The star is active and variable at the percent level \citep{Pont2007p1347}, and solar flares have been known to produce variability in the aurorae of giant planets in our own Solar System \citep{pryor2005p312}.  We cannot rule out this possibility, but we believe that any flaring event strong enough to produce the emission feature seen by S10 would produce enough high-energy radiation to dissociate the emitting molecules; additionally, there would be many other regions of the spectrum that would show significant variability. A second scenario would be a decrease in the number of emitting molecules in the planet's atmosphere, possibly due to an increase in cloud opacity; however, no variability was detected at 8\,$\mu$m by \citet{Agol2010p1861} over seven secondary eclipses between 2006 and 2008 down to a level of 2.7\%, suggesting that cloud variability over these timescales is minimal. 

A third source of variability which could be thought to mimic the effects of a planetary eclipse would be a change in the stellar radiation through the rotational modulation of stellar spots, which can add time-varying color-dependent effects \citep{Pont2008p109}, or a darkening of the planet's emission due to features on the planetary limb. However, we expect any modulations of the stellar or planetary brightness within a single night to be far below the level needed to produce the features seen by S10 (especially given that the star's rotation period is 11 days \citep{Winn2007p1828}), and the variability would almost certainly not be restricted to a single spectral channel.

\subsection{Possible Broad Emission Features}
The absence of narrow emission lines in our high-resolution residual spectrum cannot rule out the presence of very broad emission features whose flux is spread over a large number of resolving elements in our spectra. The results presented by S10 are grouped into 0.1\,$\mu$m spectral channels, while our {\it entire} spectrum covers only a fraction of one bin ($\sim0.05\,\mu$m) at high resolution; therefore broad underlying features would be removed by continuum fitting in our reduction process. However, the primary sources of broad spectral features - broadened gas-phase lines and PAH emission - can be rejected. Broadening of individual molecular lines would need to approach 130 km/s for this explanation to be viable, requiring either extreme pressure broadening or opacity broadening. Based on broadening coefficients \citep{pine1992p773}, about 20 atmospheres of pressure would be needed; these high densities and pressures are incompatible with NLTE emission, so pressure broadening can be excluded. Opacity broadening could result from extreme saturation of methane or water lines near 3.3\,$\mu$m, similar to those commonly seen in transmittance spectra of Earth's atmosphere (e.g., Fig. \ref{example}, near 3.280\,$\mu$m), but would imply similarly bright emission at 2\,$\mu$m and at 6\,$\mu$m (water) or 8\,$\mu$m (methane); this emission has not been detected in earlier HST \citep{swain2009pL114} and Spitzer/IRS \citep{grillmair2008p767} observations, and the K-band results reported by S10 also show no such emission. Alternately, the  emission feature could be intrinsically broad; in particular, a broad emission band at 3.3\,$\mu$m due to polycyclic aromatic hydrocarbons (PAHs) has been detected in a wide range of astrophysical environments \citep{allamandola1989p733}.  However, in this scenario PAH band emission would also be expected at longer wavelengths (6.2\,$\mu$m, 7.7\,$\mu$m, 8.7\,$\mu$m, and 11.3\,$\mu$m), with similar or even brighter intensity than the 3.3\,$\mu$m feature. These features are not detected in the Spitzer/IRS results. 

\begin{figure*}[htb]
\centering{
\includegraphics[width=170mm]{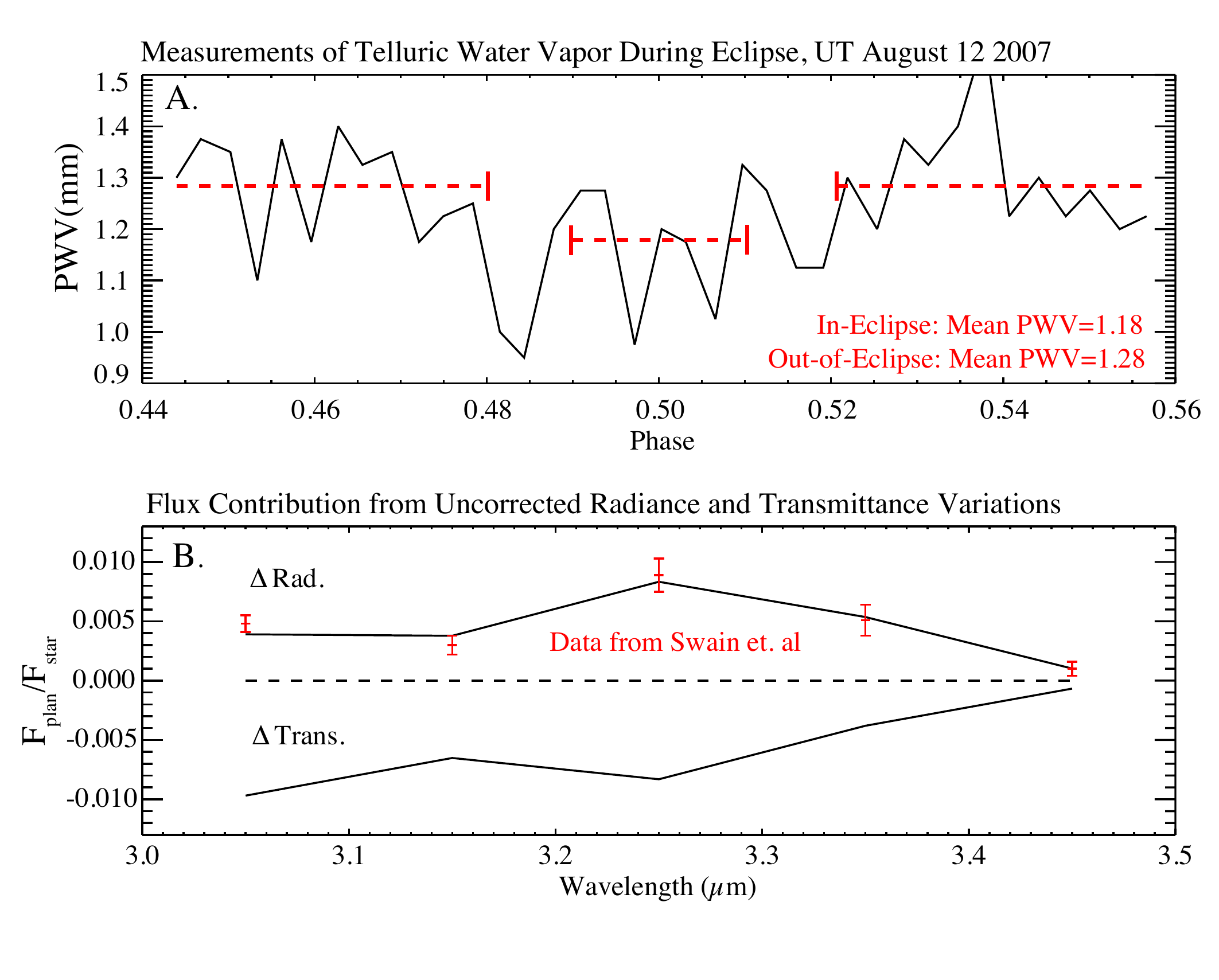}
\caption{\label{telluric} A plot of the effects of changes in telluric water absorption during the observing night analyzed by \citet{swain2010p637} (UT August 12, 2007).  Telluric water vapor affects both the transmittance (how much star light reaches the detector) and the sky radiance (how much background thermal emission arrives from the atmosphere). A) Measurements of precipitable water vapor (PWV) using data from the tipping radiometer at the CSO on Mauna Kea (black), with the mean in-eclipse and out-of-eclipse values overplotted (red). B) We calculated the changes in transmittance ($\Delta$ Trans.) and radiance ($\Delta$ Rad) for the measured change in water vapor, and compared them to the results from \citet{swain2010p637}.  While we cannot identify with certainty the cause of the differences between our data and the results from \citet{swain2010p637}, the similarity between the effects of changes in sky radiance and the Swain et al. results is consistent with inaccurate removal of variable telluric features.  }
}
\end{figure*}

\subsection{Telluric Contamination}
\label{cont} The final explanation we consider is that the excess flux reported by S10 may have been caused by systematic contamination due to imperfect removal of variable telluric absorption or emission. At L-band wavelengths, one of the largest sources of variability in flux over timescales longer than 15 minutes is the change in atmospheric conditions.  Molecular species in the Earth's atmosphere affect observations in the L band in two ways: by reducing the stellar flux transmitted through the atmosphere (transmittance) and by emitting thermal radiation with a Boltzmann temperature characteristic of the lower atmospheric layers (radiance).  Radiance is an additive component and must be subtracted from the stellar flux (usually accomplished by using an ABBA nodding sequence), while transmittance is corrected by dividing by a normalization factor based on a determination of the amount of flux lost (in our data reduction process, this is accomplished using models of telluric extinction).

Several lines of evidence buttress the the hypothesis that the S10 result is due to inadequate removal of telluric contamination.  First, the best-fitting molecular model in all L-band spectral channels presented by S10 is an emission model of water at 300K (see Figure \ref{models}), but we see no evidence of velocity-shifted emission lines in our data.  While the characteristic temperature of the atmosphere of HD\,189733b would be expected to be closer to 1200K, the characteristic temperature for water in our own atmosphere is 296K. This suggests that a change in terrestrial water vapor during the planetary eclipse could be the cause of the observed variation in eclipse depth with wavelength.

S10 analyzed data for a single eclipse observed on UT August 12, 2007 (M. Swain, private communication).  Measurements of submillimeter opacity can be used as a proxy for precipitable water vapor (PWV; \citet{masson1994p87}), and measurements from the tipping radiometer on Mauna Kea \citep{radford2002p148} indicate that the average PWV was significantly higher ($\sim9\%$) in the hours before and after the eclipse, compared with the period of the eclipse (Figure \ref{telluric}A).  This variation would result in a higher telluric transmittance during eclipse (contributing to a decrease in the uncorrected eclipse depth), but would also result in a lower telluric radiance (contributing to an increase in the uncorrected eclipse depth). Using the LBLRTM atmospheric modeling code, we have generated telluric transmittance and radiance spectra for two water vapor abundances - the average PWV during eclipse (1.18 mm) and the average PWV before and after eclipse (1.28 mm). To convert the radiance variation to a total flux we must multiply by the slit area; we use a square with dimensions of the slit width used by S10 (1.6 arcseconds).  In Figure \ref{telluric}B, we show the change in the telluric radiance emission in units of the stellar flux ($\Delta$Rad) and the change in stellar flux due to extinction by the telluric transmittance function ($\Delta$Tran). The fluxes shown for each spectral channel are the contributions from each effect on the apparent eclipse depth in that channel, if not compensated for in the original analysis by S10. 

The similarity between the signal observed by S10 and the contribution from the sky radiance due to changes in the telluric water vapor emission suggests that the effects of atmospheric transmittance may have been removed accurately (possibly by normalizing each spectrum by the mean baseline flux) but that the effects of telluric radiance may have been incompletely corrected, possibly because S10 processed their A and B beams separately. Incorrect compensation for the effects of sky radiance would lead to an apparent drop in flux during the period of the eclipse and result in a spurious measurement for the loss of flux from the planet. This effect may have been exacerbated by the wide slit width used, which would have made accurate removal of the sky emission more difficult.  The apparent ability of the authors to match the previous HST observations at K-band wavelengths further supports radiant sky emission as opposed to a problem with transmittance correction, since the intensity of telluric radiance at K-band wavelengths is much lower than in the L band.  

Eventually it should be possible to directly assess the contribution of telluric contamination in the S10 results, but that assessment will require re-analysis of the S10 data using their original methodology, which uses non-standard analysis techniques that are described in insufficient detail in their paper to be reproduced, and comparison of these results with an analysis using independent methodology.  Although our observations and analysis unequivocally contradict the S10 results, our analysis algorithms - being optimized for high spectral resolution data - do not allow us to test the specific methodology used by S10 with equal rigor. While we cannot confirm sky emission as the source of the feature reported by S10, in the absence of a viable astrophysical explanation we consider it as the most likely possibility. 

\section{Conclusions}
We report a non-detection of emission at 3.3\,$\mu$m from the extrasolar planet HD\,189733b using high-resolution spectroscopic observations taken with the NIRSPEC instrument on the Keck-II telescope, a result in contrast with the detection of strong emission at these wavelengths announced by \citet{swain2010p637}.  We modeled the expected signal using ro-vibrational emission models of molecular species with transitions in the relevant wavelength region using a wide range of rotational excitation temperatures and vibrational level intensities, scaled to the results presented by \citet{swain2010p637}, in order to predict the line flux required at high resolution.  No flux was present at any of the expected transition frequencies between 3.27 and 3.31\,$\mu$m, with upper limits 40 times smaller than the expected line fluxes. The conditions that would lead to broad emission features beyond our detection limits are extremely difficult to reconcile with realistic models and previous observations. Additionally, our analysis indicates that the emission, if real, is too bright to be produced by fluoresence.  Our wavelength region only covers a small section of the spectrum published by S10, and we cannot rule out an emission mechanism that produces flux outside our band pass. Additionally, we cannot rule out an exotic highly time-variable stellar emission process such as charged-particle excitation due to flares.  However, we regard these explanations as improbable, and conclude that inadequate telluric correction is the most likely explanation for the \citet{swain2010p637} results.

\acknowledgements A.M.M. was supported by the Goddard Center for Astrobiology and the NASA Post-doctoral Fellowship Program. H.A.K. is supported by a fellowship from the Miller Institute for Basic Research in Science. We thank Dr. Linda Brown for directing our attention to the CH4@Titan online database. The data presented herein were obtained at the W.M. Keck Observatory, which is operated as a scientific partnership among the California Institute of Technology, the University of California and the National Aeronautics and Space Administration. The Observatory was made possible by the generous financial support of the W.M. Keck Foundation. The authors wish to recognize and acknowledge the very significant cultural role and reverence that the summit of Mauna Kea has always had within the indigenous Hawaiian community.  We are most fortunate to have the opportunity to conduct observations from this mountain. We also thank the anonymous referee for constructive comments.

%{\it Facilities:} \facility{Keck:II (NIRSPEC)}

\bibliographystyle{apj}
\bibliography{refs}

\end{document}